\documentclass[onecolumn,reprint,superscriptaddress,nofootinbib,amsmath,amssymb, aps,onecolumn,notitlepage]{revtex4-1}

\usepackage[a4paper, margin=2.50cm]{geometry}

\usepackage{bbm}
\usepackage{graphicx}
\usepackage{fancyhdr,color}
\usepackage{hyperref} 
\usepackage{mathtools}

\newcommand{\bra}[1]{\langle{#1}|}

\newcommand{\ket}[1]{|{#1}\rangle}

\newcommand{\ip}[1]{\langle{#1}\rangle}
\newcommand{\op}[2]{|{#1}\rangle\langle{#2}|}

\newcommand{\eq}[1]{Eq.~(\ref{#1})}
\newcommand{\eqs}[1]{Eqs.~(\ref{#1})}
\newcommand{\eqr}[1]{(\ref{#1})}
\newcommand{\up}{\uparrow}
\newcommand{\dn}{\downarrow}

\newcommand{\D}{{\rm d}}

\begin{document}

\fancyhead[C]{\sc \color[rgb]{0.4,0.2,0.9}{Quantum Thermodynamics book}}
\fancyhead[R]{}

\title{Information Erasure}

\author{T. Croucher}
\author{J. Wright}
\author{A.R.R. Carvalho}
\affiliation{Centre for Quantum Dynamics, 170 Kessels Road, Griffith University, Brisbane 4111, AUSTRALIA.}
\author{S.M. Barnett}
\affiliation{School of Physics and Astronomy, University of Glasgow, Kelvin Building, University Avenue, Glasgow G12 8QQ, UNITED KINGDOM}

\author{J.A. Vaccaro}
\email{J.A.Vaccaro@griffith.edu.au}
\affiliation{Centre for Quantum Dynamics, 170 Kessels Road, Griffith University, Brisbane 4111, AUSTRALIA.}

\date{\today}

\begin{abstract}

Information is central to thermodynamics, providing the grounds to the formulation of the theory in powerful abstract statistical terms.   One must not forget, however, that, as put by Landauer, {\it information is physical}. This means that the processing of information will be unavoidably linked to the costs of manipulating the real physical systems carrying the information. Here we will focus on the particular process of erasing information, which plays a fundamental role in the description of heat engines. We will review  Landauer's principle and the associated erasure energy cost. We will also show, following the recent contributions from Vaccaro and Barnett, that cost of erasing does not need to be paid with energy, but with any other conserved quantity. Finally, we will address the issue of designing heat engines based on these new concepts.
\end{abstract}

\maketitle

\thispagestyle{fancy}

\section{Introduction} \label{sec:intro}

The origin of thermodynamics is intimately related to the development of heat engines in the $18^{th}$ century. Understanding the connections between macroscopic quantities such as pressure, volume and temperature, as well as the interplay between work and heat in engine cycles, led to the formulation of the fundamental laws underpinning thermodynamical transformations. The analysis of the Carnot heat engine, for example, is the root behind the second law of thermodynamics. 

The thermodynamical behaviour of macroscopic systems can also be understood as emerging from microscopic dynamical laws and the intrinsic statistical uncertainty about the state of the system. This statistical approach to thermodynamics, pioneered by Maxwell, Boltzmann, and Gibbs, brought to centre stage the concept of information, which is our main concern in this chapter. Perhaps the first instance where information and thermodynamics are linked together is in the scenario of Maxwell's demon, where a hypothetical microscopic intelligent being is able to follow the state of individual molecules and act on the system to apparently violate the second law. 

While Maxwell himself~\cite{Maxwell1902} seemed to be open to the possibility that our conclusions about the behaviour of macroscopic systems may simply not be applicable to situations where individual constituents can be observed and manipulated\footnote{Maxwell states in his book~\cite{Maxwell1902}: ``This is only one of the instances in which conclusions which we have drawn from our experience of bodies consisting of an immense number of molecules may be found not to be applicable to the more delicate observations and experiments which we may suppose made by one who can perceive and handle the individual molecules.''}, most of the subsequent work on Maxwell's demon paradox focused on trying to reconcile the second law with the underlying dynamical microscopic description of the system \cite{Leff1990,Leff2003}. 

An important contribution in this direction was made by Szilard~\cite{Szilard1929}, who designed a single molecule heat engine that would work in apparent violation of the second law, provided that information about the partition occupied by the molecule were measured and recorded. Szilard circumvents this violation by asserting that such acquisition of information would necessarily incur on an entropy increase not less than $k_B \ln{(2)}$, although he doesn't pinpoint when such an increase occurs in his engine cycle. Subsequently, Brillouin~\cite{Brillouin1953,Brillouin1965} showed that the same entropy bound follows from the act of measurement itself, narrowing down the possible source of entropy cost.

A few years later, in principle in an unrelated context, Landauer~\cite{Landauer1961} investigated the question of the physical cost of performing basic computation. Using the simplest irreversible logical operation of resetting an unknown bit of information, Landauer showed that this {\it erasure} process would require a minimum of $k_{B} T \ln{(2)}$ of energy to be dissipated as heat into the environment surrounding the bit. But it was only twenty years that later that Bennett~\cite{Bennett1982,Bennett1987} established the connection between Landauer's erasure principle and  the Maxwell's demon problem. Moreover, Bennett also proposed a reversible measurement scheme that wouldn't involve thermodynamical costs, proving that the real reason why the demon doesn't violate the second law is the necessity to erase the demon's memory rather than any energy costs associated with measuring and storing the information.      

Formulating the problem in information-theoretical terms allows one to detach the statistical aspects of the theory from the specificities of any physical realization, a view that is at the core of Jaynes' maximum entropy principle~\cite{Jaynes1957,Jaynes1957a}. Using statistical arguments, Jaynes argues that one could move away from a standpoint where energy plays a privileged role in the theory and treat different conserved quantities on equal footing. 

But not until 2006 was the potential of Jaynes' generalisation of statistical mechanics for erasure and heat engines appreciated~\cite{Vaccaro2006,Vaccaro2009a}. Vaccaro and Barnett realised that under Jaynes' framework, it was natural to question the special role of energy in Landauer's erasure mechanism. In their work, for example, they propose an erasure that works at the cost of angular momentum rather than energy. In fact, any physical resource that could account for the increase in entropy in the reservoir as it absorbs the entropy from the erased information would be an equally valid choice. 

This brings us back to where we started: the cornerstone of thermodynamics, the heat engine. Since Landauer's principle explains that an erasure step is required to allow Szilard's engine to work without violating the second law, the ideas put forward in~\cite{Vaccaro2006,Vaccaro2009a} indicate that heat engines can work with a single thermal reservoir, as long as there is another process that takes care of the erasure as, for example, a reservoir of another conserved quantity.

This chapter is organised as follows. In the next section, we discuss in more detail the principles of information erasure, from Landauer's energetic cost to Vaccaro-Barnett's approach in terms of multiple conserved quantities. Then, we describe how these ideas can be used to design fundamentally different heat engines. Finally, we discuss fluctuation theorems associated with erasure and end with a discussion.


\section{The general principles of information erasure}\label{sec:general principles}

We should be clear from the outset what is meant by erasing information.
In particular, we need to distinguish between \emph{losing} knowledge about a system and \emph{resetting} its state.
For example, the state of an open system may evolve from a pure state to a mixed state by interacting with its environment.
As the pure state represents complete knowledge of the system and the mixed state represents incomplete knowledge, our knowledge of the system is lost over time.
The extreme case where the state becomes maximally mixed represents a complete loss of knowledge.
However, to say that information has been erased in this case would not be following conventional terminology, although this use is occasionally found in the literature \cite{Apollaro2014}.
Rather, the term erasure is more-often reserved for the quite distinct task of resetting the state of a system.
An example of setting a state is given by the state preparation that typically occurs at the beginning of an experiment in many branches of quantum physics.
Resetting would occur if the same system is reused for a subsequent experiment.
Landauer was interested in the task of resetting the memory systems of computing machines.
For each particular computation performed by the machine, the state of the memory system encodes a corresponding result and, in this context, the state represents information.
However, the resetting task needs to be independent of any particular computation, and so must be capable of resetting a memory system that contains an arbitrary result.
As an arbitrary result is represented by the maximally mixed state, the objective of resetting is to replace a maximally mixed state with a pure reset state.
This is the conceptual framework of \emph{erasure of information}: the arbitrary information that is encoded in the memory system is erased by physically manipulating the memory system so that it evolves from the associated maximally mixed state into the pure reset state.

To explore how information erasure can be performed, let the memory system be a two state system whose state space is spanned by the basis $\{\ket{0},\ket{1}\}$ labelled according to corresponding binary values, and let the reset state be $\ket{0}$.
The erasure process needs to replace the maximally mixed state $\frac{1}{2}(\op{0}{0}+\op{1}{1})$ with the pure state $\op{0}{0}$.
This requires the two orthogonal states $\ket{0}$ and $\ket{1}$ to be mapped to the same state $\ket{0}$, which rules out a unitary operation on the memory system alone.
It can, however, be represented by a unitary SWAP operation on the combination of the memory and an ancillary system that has been prepared in the reset state $\ket{0}$, i.e.
\begin{align}
     \frac{1}{2}(\op{0}{0}+\op{1}{1})_{M}\otimes\op{0}{0}_{A}
     \xrightarrow{\text{SWAP}}\op{0}{0}_M\otimes\frac{1}{2}(\op{0}{0}+\op{1}{1})_A
\end{align}
where $M$ and $A$ label the memory and ancilla density operators, respectively.
Of course, this immediately raises the question of the preparation of the state of the ancilla.
According to Jaynes' maximum entropy (MaxEnt) principle, without knowing anything about the history of the ancilla, the best state that describes it is a maximally mixed one.
The situation of ancilla being freely available in the reset state $\ket{0}$ is, therefore, an unlikely one and so using a SWAP operation merely defers the problem of erasure to the preparation of the ancilla system.
To avoid this problem, we need to consider erasure process that make use of freely available resources.

Thermal reservoirs at a fixed temperature $T$, such as the temperature of the local environment, are regarded as freely available \cite{Brandao2015}[Ch 25].
As such, the task of erasure needs to be cast in terms of transferring the entropy from the memory system into a thermal reservoir at finite temperature $T$.
This is essentially the problem that Landauer considered \cite{Landauer1961}.

If the erasure is to be most efficient it should be reversible and, fortunately, this makes calculating the minimal physical effort of an erasure process easy to quantify.
As the total entropy of the memory and reservoir system is fixed for a reversible process, any change of $\Delta S_M$ in the entropy of the memory is accompanied by an opposing change of $\Delta S_R=-\Delta S_M$ in the entropy of the reservoir.
If the reservoir is sufficiently large that its temperature $T$ remains constant throughout the erasure, it will absorb the corresponding amount $Q_R=T\Delta S_R$ of heat in the process.
The 1 bit of information that is erased from the memory represents a change in Gibbs entropy of $\Delta S_M = -k_B\ln(2)$, where $k_B$ is Boltzmann's constant, and so $Q_R=k_B T\ln(2)$.
This is Landauer's result that the minimal cost of the erasure of 1 bit of information is the dissipation of heat of $k_B T\ln(2)$ in the surroundings \cite{Landauer1961}.
Landauer did not prescribe any particular method for the erasure process, rather he asserted only that a minimum heating effect accompanies the erasure of information in general, and gave the example of heat of $k_BT\ln(2)$ per bit being transferred to the surroundings under equilibrium conditions \cite{Landauer1961}.  This general effect is now known as Landauer's erasure principle.

It is instructive to have a simple dynamical model of an erasure process, and one that serves this purpose well involves doing work on the memory system to slowly increase the gap in the energy of its states $\ket{0}$ and $\ket{1}$ while keeping it in thermal equilibrium with the reservoir at fixed temperature $T$.
Let the energy gap $E$ be produced by increasing the energy of the state $\ket{1}$ by a small amount, $\D E$, while keeping the energy of the state $\ket{0}$ unchanged.
As the memory system is maintained in thermal equilibrium with the reservoir, its state when the energy gap is $E$ is described by the density operator
\begin{align}
   \hat{\rho} = \frac{\ket{0}\bra{0}+e^{-E/k_BT}\ket{1}\bra{1}}{1+e^{-E/k_BT}}\ .
\end{align}
The work required to increase the gap from $E$ to $E+\D E$ is given by
the probability, $p_1$, of the occupation of the state $\ket{1}$ multiplied by $\D E$, that is $\D W=p_1 \D E$, where $p_1=e^{-E/k_BT}(1+e^{-E/k_BT})^{-1}$.
The total work done in increasing the gap from zero to infinity\footnote{In practical terms, the fact that the probability $p_1$ reduces exponentially with $E$, means the work process may be halted at a finite value of the gap for a correspondingly small probability of error.} is $W_M=\int \D W=k_BT\ln2$ \cite{Landauer1961}.
At the same time as the gap increases from $E$ to $E+\D E$, heat of $\D Q=E\D p_1$ is transferred between the memory and the reservoir.
Writing $E=k_BT\ln[(1-p_1)/p_1]$ and integrating over $p_1$ from $\frac{1}{2}$ to 0 gives the total heat transferred as $Q_M=-k_BT\ln(2)$ and so heat of $Q_R=-Q_M= k_BT\ln(2)$ is transferred to the reservoir, as before.
Once the work process is completed, the memory system will be in the reset state $\ket{0}$.
It is then thermally isolated from the reservoir and the energy degeneracy of its states is restored without any further work cost or gain.
The process can be described as follows:
\begin{quotation}
\noindent
\textbf{Erasure by increasing energy gap:} \it entropy is transferred from the memory system to the reservoir, under equilibrium conditions and the conservation of energy, as the energy gap between the information-carrying states of the memory system is slowly increased.
\end{quotation}

We now use this description to generalise Landauer's erasure principle.
Landauer repeatedly justified his claim that ``information is physical'' on the basis that information is inevitably tied to physical degrees of freedom and thus to physical laws \cite{Landauer1993}.
An implicit consequence of this is that there is a minimal \emph{physical cost} associated with the erasure of information.
Although he regarded the transfer of entropy from the memory system to the surroundings as an essential part of erasure, he demonstrated the physical cost only in terms of \emph{heat and energy} \cite{Landauer1961,Landauer1993}.
In our description of a specific erasure process, energy appears in two important contexts: in the physical cost of increasing the energy gap, and as a conserved quantity.
If the conserved quantity was something other than energy, then the cost of the erasure process might also be in terms of the different quantity and not energy.
It was a rumination of this kind that led to our generalisation of Landauer's principle \cite{Vaccaro2006,Vaccaro2009a,Vaccaro2011,Barnett2013}.
Taking account of Jaynes' generalisation of Gibbs ensemble for the case where the state of the reservoir involves multiple physical observables $\hat{V}_k$ for $k=1,2,\ldots$ allows us to find the general cost of erasing information as follows \cite{Barnett2013}.
The role played by the $k$-th observable $\hat{V}_k$ is like that of energy in conventional thermodynamics and, following Jaynes \cite{Jaynes1957}, will be referred to as the $k$-th type of energy.
The density operator giving the best description of the reservoir when the averages $\ip{\hat{V}_k}$ are known is \cite{Jaynes1957}
\begin{align}
   \hat \rho = \exp\Big(-\mu-\sum_k\lambda_k\hat V_k\Big)
\end{align}
and its Shannon entropy is
\begin{align}  \label{eq:S for general reservoir}
    S = \mu+\sum_k\lambda_k\ip{\hat V_k}
\end{align}
where $\mu$ and $\lambda_k$ are Lagrange multipliers.
We use the Shannon entropy \cite{Shannon1963} here because it is dimensionless and unbiased with respect to the quantities $\hat{V}_k$, in contrast to the Gibbs entropy which has dimension of energy per Kelvin.
Allowing changes in $\hat V_k$ to be independent of those of the state of the reservoir implies that the corresponding changes $\ip{\delta\hat{V}_k}$ and $\delta\ip{\hat{V}_k}$ are independent, and so \cite{Jaynes1957}
\begin{align}  \label{eq:delta S}
        \delta S=\sum_k\lambda_k\delta Q_k
\end{align}
where
\begin{align}  \label{eq:Q_k}
        \delta Q_k\equiv \delta\ip{\hat{V}_k}-\ip{\delta\hat{V}_k}
\end{align}
corresponds to the $k$-th type of heat.
Rearranging \eq{eq:Q_k} as
\begin{align}  \label{eq:1st law for k}
        \delta U_k=\delta W_k+\delta Q_k
\end{align}
expresses the first law for $\hat{V}_k$ in terms of the change $\delta U_k\equiv\delta\ip{\hat{V}_k}$ in the $k$-th type of internal energy, the amount  $\delta W_k\equiv\ip{\delta\hat{V}_k}$ of the $k$-th type of work done on the reservoir, and the amount $\delta Q_k$ of the $k$-th type of heat transferred to it.

If this more general reservoir is used for the erasure of 1 bit of information in a reversible, and thus efficient, manner then the total entropy of the reservoir and memory system will be unchanged.
Allowing for irreversible processes, the information erased from the memory appears as an increase of $\Delta S\ge\ln(2)$ nats in the entropy of the reservoirs and, from \eq{eq:delta S}, this implies
\begin{align} \label{eq:general cost of erasure of 1 bit}
        \sum_k\lambda_k\Delta Q_k \ge \ln(2)
\end{align}
which was first derived in Ref.~\cite{Barnett2013}.
This result represents the reservoir gaining an entropy of at least $\ln(2)$ nats by the transfer of different types of heat $\Delta Q_k$ to the reservoir with the only restriction being that their $\lambda_k$-weighted sum is bounded below by $\ln(2)$.
There are a number of comments to make about this result.
The first is that if the cost is paid in only one type of heat, say for $k=1$, it can be expressed as
\begin{align}  \label{eq:erasure cost Q_1}
       \Delta Q_1\ge \frac{\ln(2)}{\lambda_1}\ .
\end{align}
In the case where $\hat V_1$ is the operator for energy $\hat H_R$, the Lagrange multiplier $\lambda_1$ is the inverse temperature $\beta=1/k_BT$, the change $\Delta Q_1$ is the heat $Q$, and we recover Landauer's result $Q=k_BT\ln(2)$ \cite{Landauer1961}.
Alternatively, in the case where $\hat V_1$ is the $z$ component of spin angular momentum $\hat{J}_z$, the Lagrange multiplier $\lambda_1$ is the corresponding inverse spin temperature $\gamma$, the change $\Delta Q_1$ is the spin-equivalent of heat $\mathcal{Q}_s$ called \emph{spintherm} \cite{Croucher2017}, where
\begin{align}   \label{eq:spintherm}
        \mathcal{Q}_s\equiv\delta\ip{\hat{J}_z}-\ip{\delta\hat{J}_z}\ ,
\end{align}
and we recover our previous result \cite{Vaccaro2009b,Barnett2013}
\begin{align}  \label{eq:spin erasure}
        \mathcal{Q}_s\ge\gamma^{-1}\ln(2)\ .
\end{align}
Lostaglio \emph{et al}. \cite{Lostaglio2017} have shown how the cost of erasing 1 bit of information can be paid in varying amounts of heat $Q$ and spintherm $\mathcal{Q}_s$ satisfying $\ln(2)=\beta Q + \gamma \mathcal{Q}_s$.

The description given above for a specific erasure method can easily be generalised to accommodate this versatility in the cost of erasure as follows:
\begin{quotation}
\noindent\textbf{Erasure by increasing arbitrary eigenvalue gap:} \it
entropy is transferred from the memory system to the reservoir, under equilibrium conditions and the conservation of the set of observables $\{\hat{V}_k:k=1,2,\ldots\}$, as the gap between the eigenvalues associated with the information-carrying eigenstates of an observable $\hat{V}_j$ of the memory system is slowly increased.
\end{quotation}
The erasure process is additive in the sense that it can be applied in a sequence of stages, each one erasing successively more information.
The choice for the index $j$ and the amount of information erased in each stage determines the proportion of the corresponding cost $\Delta Q_j$ in \eq{eq:general cost of erasure of 1 bit} \cite{Lostaglio2017}.


\section{Erasure of information carried by discrete variables}\label{sec:discrete variables}

The descriptions of the specific erasure processes in the preceding section assume that the eigenvalues corresponding to the information-carrying eigenstates of the observable $\hat{V}_j$ can be varied continuously.
In the particular case of energy, although the information-carrying states may belong to a discrete portion of the eigenvalue spectrum, the spacing of the spectrum can be adjusted, in principle, through the use of an external potential.
However, observables such as angular momentum do not have this possibility for fundamental reasons and so they call for special treatment.
In this section, we  use the $z$ component of spin angular momentum, $\hat{J}_z$, as our prototypical discrete observable to illustrate the kinds of issues that arise.

Although it is not possible to vary the eigenvalues of a spin observable, an increase in the gap between the information-carrying eigenstates can be \emph{simulated} as follows.
Consider a memory system composed of a spin-$\frac{1}{2}$ particle where $\hat{V}_j$ corresponds to the $z$ component of angular momentum  and the information-carrying states are its eigenstates.
By augmenting the memory spin with a collection of ancilla spin-$\frac{1}{2}$ particles, it is possible to select pairs of eigenstates of the $z$ component of total angular momentum whose eigenvalues differ by $\Delta J_z=n\hbar$ for $n=1,2,\ldots$.
A sequence of pairs with increasing differences, i.e. $\Delta J_z=\hbar, 2\hbar,\ldots$, can be used as the information-carrying states as described in Refs.~\cite{Vaccaro2011,Barnett2013}.
The operation of transferring the information from one pair to the next in the sequence requires the corresponding $k$-th type of work, i.e.
\begin{align}   \label{eq:spinlabor}
        \mathcal{L}_s\equiv \ip{\delta\hat{J}_z}\ ,
\end{align}
which has been called \emph{spinlabor} \cite{Croucher2017}.
However, as the increase in the gap is not continuous but rather in steps of $\hbar$, the simulation is only approximate.
Nevertheless, the erasure process will be approximately quasi-static provided this step size is small compared to the spin temperature, i.e. $\hbar\ll\gamma^{-1}$.
Being not quite reversible, the actual cost of erasure in this case would be higher than the lower bound in \eq{eq:spin erasure}.

Another issue that arises when treating a spin-$\frac{1}{2}$ memory system is that the memory requires angular momentum of $\frac{1}{2}\hbar$, on average, to go from the initial maximally-mixed state,\footnote{We use subscripted labels $M$ and $R$ to distinguish quantities associated with the memory and reservoir, respectively, when confusion might otherwise arise.}
\begin{align}  \label{eq:mixed state}
        \frac{1}{2}(\op{\up}{\up}+\op{\dn}{\dn})_{M}\ ,
\end{align}
to the reset state $\op{\up}{\up}_M$, where $\ket{\dn}$, $\ket{\up}$ are the eigenstates of $\hat{J}_z$ corresponding to the eigenvalues $-\frac{1}{2}\hbar$, $\frac{1}{2}\hbar$, respectively.
This increases the amount of spinlabor needed as follows.
The first law for $\hat{J}_z$  is given, according to \eq{eq:1st law for k} with the definitions \eqs{eq:spintherm} and \eqr{eq:spinlabor}, by
\begin{align}
        \delta\ip{\hat{J}_z}_R =\mathcal{L}_{s,R}+\mathcal{Q}_{s,R}\ ,\qquad
        \delta\ip{\hat{J}_z}_{MR} =\mathcal{L}_{s,MR}+\mathcal{Q}_{s,MR}\ ,
\end{align}
for the reservoir (labeled $R$) and the memory-reservoir combined system (labeled $MR$), respectively.
Taking note of the facts that spin angular momentum and spinlabor are additive, i.e. $\delta\ip{\hat{J}_z}_{MR}=\delta\ip{\hat{J}_z}_M+\delta\ip{\hat{J}_z}_R$ and $\mathcal{L}_{s,MR}=\mathcal{L}_{s,M}+\mathcal{L}_{s,R}$, respectively, no spinlabour is performed on the reservoir, i.e. $\mathcal{L}_{s,R}=0$, the erasure produces the change $\delta\ip{\hat{J}_z}_M=\frac{1}{2}\hbar$ in the memory, and no spintherm is exchanged between the memory-reservoir combined system and its environment, i.e. $\mathcal{Q}_{s,MR}=0$, reveals \cite{Croucher2017}
\begin{align}  \label{eq:less hbar of spintherm}
        \mathcal{Q}_{s,R}=\mathcal{L}_{s,M}-\mbox{$\frac{1}{2}$}\hbar\ .
\end{align}
This shows that $\frac{1}{2}\hbar$ of the spinlabor $\mathcal{L}_{s,M}$ is retained by memory and the remainder is dissipated as spintherm in the reservoir.
If, instead, the reset state was $\op{\dn}{\dn}_M$, the result would be
\begin{align} \label{eq:extra hbar of spintherm}
        \mathcal{Q}_{s,R}=\mathcal{L}_{s,M}+\mbox{$\frac{1}{2}$}\hbar\ ,
\end{align}
i.e. there would be an additional contribution of $\frac{1}{2}\hbar$ to the spintherm of the reservoir due the change in the state of the memory.
The same value of $\mathcal{Q}_s$ occurs in \eqs{eq:less hbar of spintherm} and \eqref{eq:extra hbar of spintherm}---it is the spintherm associated with a transfer of $\ln(2)$ in entropy to the reservoir and it is bounded below by \eq{eq:spin erasure}.
The values of the spinlabor $\mathcal{L}_{s,M}$ in \eqs{eq:less hbar of spintherm} and \eqref{eq:extra hbar of spintherm} are, however, different and reflect the physical effort needed for different tasks. 

\begin{figure}[htb]
\centering
  \includegraphics[width=0.62\textwidth]{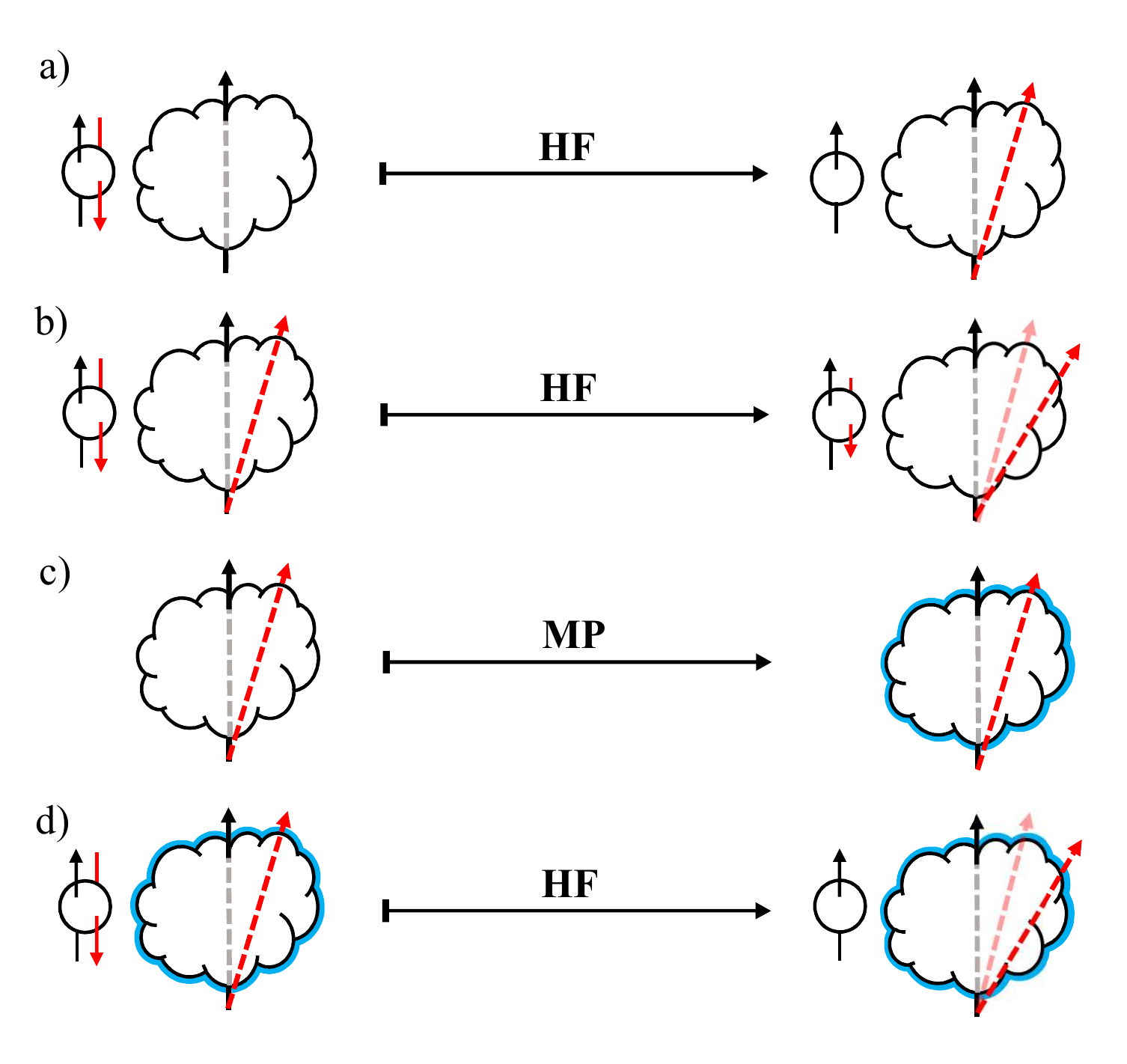}
  \caption{\label{fig:pulse_evolution} 
  Sketches representing the mappings for the partial fixed-point erasure process. The circle represents the memory spin and the cloud represents the spin reservoir.  Arrows represent the direction of spin angular momentum and their colours represent different states in a statistical mixture. The effect of the magnetic pulse on the reservoir is represented by a blue halo. Panel (a) represents the mapping induced by the hyperfine interaction in \eq{eq:HF mapping}  for erasing the first bit of information, (b) represents the incomplete erasure when the same mapping is applied to the second bit, (c) represents the mapping induced by a spatially-varying magnetic pulse in \eq{eq:MP mapping}, and (d) represents the complete erasure of the second bit following the magnetic pulse.  }
\end{figure}

Finally, discrete observables also allow different kinds of erasure mechanisms.
We briefly describe here one that arises within the framework of the central spin problem; its full details may be found in Ref.~\cite{Wright2018}.
The memory is represented by the central spin, typically an electron, and the reservoir is a collection of spins, typically nuclei, that surround it.
We assume the hyperfine interaction between the electron and nuclei is the dominant interaction and that other interactions can be ignored.
Let the reservoir comprise spin-$\frac{1}{2}$ particles that are initially in the fully polarised spin-up state $\ket{0}_R\equiv \ket{\uparrow\uparrow\uparrow\cdots}_R$ where the $0$ represents no particle is spin down.
In general, we will write $\ket{n}_R$ to represent a future-evolved state of the reservoir that is a sum of different permutations of $n$ nuclei spin down and the remainder spin up.
The memory is initially in the mixed state given by \eq{eq:mixed state} and the hyperfine interaction induces the memory and reservoir to exchange spin angular momentum at a particular rate, flipping the spin of the memory in a way that conserves total angular momentum.
Note however, that the state $\ket{\up}_M\ket{0}_R$ is a fixed point of the evolution because exchanging spin is not possible, whereas the state $\ket{\dn}_M\ket{0}_R$ evolves to $\ket{\up}_M\ket{1}_R$ at half the spin flopping period.
This situation is represented by the mapping\footnote{For brevity, we ignore irrelevant phase factors when writing down states of the memory and reservoir.}
\begin{align}
        \ket{\up}_M\ket{0}_R & \xmapsto{\ \mbox{\tiny HF}\ } \ket{\up}_M\ket{0}_R  \quad\mbox{(fixed point)}\\
        \ket{\dn}_M\ket{0}_R & \xmapsto{\ \mbox{\tiny HF}\ } \ket{\up}_M\ket{1}_R\ . 
\end{align}
Applying the mapping to the initial state of the memory and reservoir gives
\begin{align}\label{eq:HF mapping}
        \frac{1}{2}(\op{\up}{\up}+\op{\dn}{\dn})_{M}\otimes\op{0}{0}_R & \xmapsto{\ \mbox{\tiny HF}\ } \op{\up}{\up}_M\otimes\frac{1}{2}(\op{0}{0}+\op{1}{1})_{R} \ .
\end{align}
which shows that entropy is transferred from the memory to the reservoir.
The memory is therefore erased and the reservoir has lost some of its polarisation as illustrated in Fig.~\ref{fig:pulse_evolution}(a).  The principle underlying this kind of erasure is that the reset state of the memory, i.e. $\ket{\up}_M$, is a fixed point of the evolution for the reservoir state $\ket{0}_R$ whereas the other memory state $\ket{\dn}_M$ is not.

Attempting to repeat the erasure a second time, however, faces the problem that the reset state $\ket{\up}_M$ is not a fixed point of the evolution for the reservoir state $\ket{1}_R$ and so an error would result as illustrated in Fig.~\ref{fig:pulse_evolution}(b).
There is a 25\% chance that this situation will occur.
Fortunately, there is a way to circumvent this problem \cite{Wright2018}.
Applying a spatially-varying, pulsed magnetic field orientated along the $z$ direction will induce relative phase shifts between the nuclei depending on their position and their spin state.
This induces the mapping, illustrated in Fig.~\ref{fig:pulse_evolution}(c),
\begin{align}
        \ket{n}_R & \xmapsto{\ \mbox{\tiny MP}\ } \ket{n,\bf{t}}_{R} 
\end{align}
where the label $\bf{t}$ characterises the pulse.
An appropriately designed magnetic pulse will have the effect that $\ket{\up}_M\ket{n,\bf{t}}_R$ is essentially a fixed point of the hyperfine evolution as illustrated in Fig.~\ref{fig:pulse_evolution}(d).\footnote{The fixed point condition is not satisfied exactly, however, it can be approached with small error, in principle.}  The combination of the hyperfine interaction and appropriate magnetic pulses gives the general mappings:
\begin{align}  \label{eq:MP+HF mapping}
        \frac{1}{2}(\op{\up}{\up}+\op{\dn}{\dn})_{M}\otimes\op{n,\bf{t}}{n,\bf{t}}_R & \xmapsto{\ \mbox{\tiny HF}\ } \op{\up}{\up}_M\otimes\frac{1}{2}(\op{n,\bf{t}}{n,\bf{t}}+\op{n+1,\bf{t}}{n+1,\bf{t}})_{R} \\
        \ket{n,\bf{t}}_R & \xmapsto{\ \mbox{\tiny MP}\ } \ket{n,\bf{t}'}_{R}\ .   \label{eq:MP mapping}    
\end{align}
The erasure process can, therefore, be operated multiple times in cycles involving the hyperfine interaction for an appropriate duration followed by a brief magnetic pulse.
It can be described as follows:
\begin{quotation}
\noindent\textbf{Erasure by partial fixed point evolution:} \it
entropy is transferred from the memory system to the reservoir by ensuring the reset state of the memory is a fixed point of the memory-reservoir interaction, and the unwanted state of the memory evolves into the reset state.
\end{quotation}
In the following section we show how it can be used in a heat engine to erase information in the working fluid.

\section{Connection with heat engines}\label{sec:heat engines}

\begin{figure}[htb]
\centering 
  \includegraphics[width=0.45\textwidth]{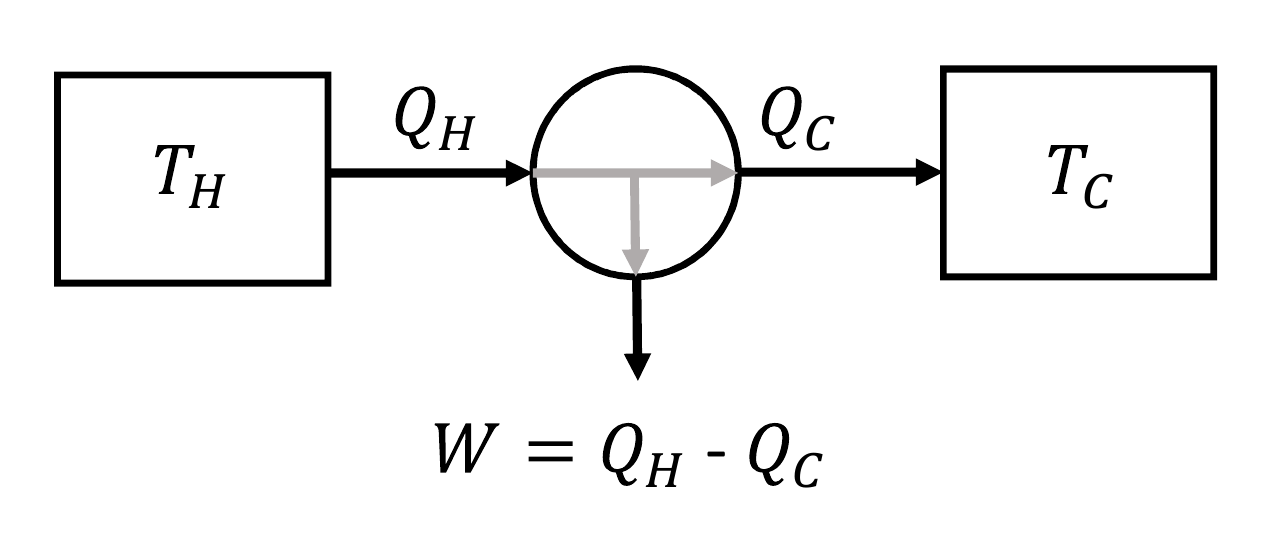}
  \caption{A Carnot engine diagram: The heat $Q_{H}$ flows from a hot body of temperature $T_{H}$ to the colder body $T_{C}$ by the second law of thermodynamics.  Not all the heat flows to the cold body, any excess is represented by mechanical or electrical work on the engine's environment.\label{fig:carnot}}
\end{figure}

As discussed in the previous sections, Landauer's erasure reconciles Maxwell's demon with the second law of thermodynamics. As a consequence, heat engines, powered by an intelligent being or any mechanical equivalent, must still work under the bounds of a Carnot cycle. This means that the heat $Q_H$ extracted from a hot reservoir is only partially converted into the work $W$ as shown in Fig.~\ref{fig:carnot}. Under Landauer's principle, the difference $Q_H-W$ represents the internal work that is needed to erase the demon's memory and be dissipated into the cold reservoir as waste heat $Q_C$.   

\begin{figure}[htb]
\centering
  \includegraphics[width=0.45\textwidth]{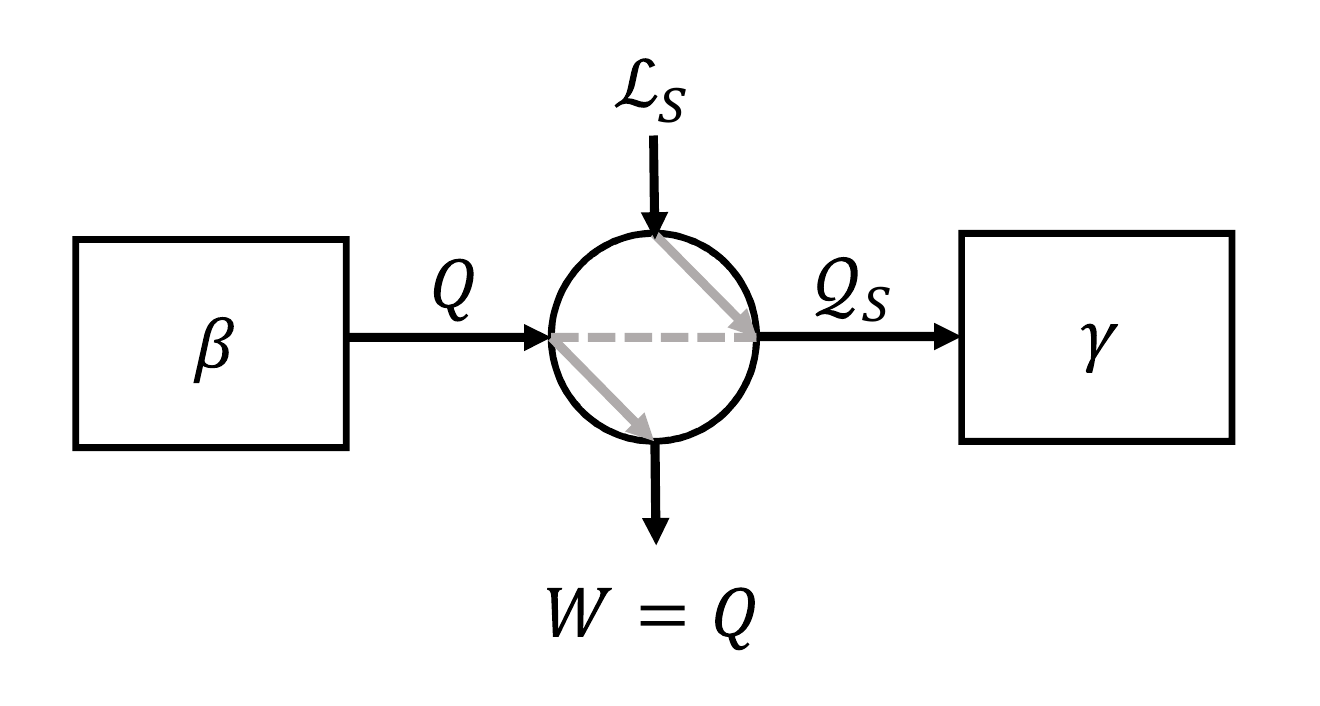}
  \caption{The conceptual diagram of the VB spin-heat engine (SHE): The work $W$ output is equal to the heat $Q$ extracted from the thermal reservoir at inverse temperature $\beta$ leaving entropy in the working fluid.  The entropy in the working fluid is erased at a cost of spinlabor $\mathcal{L}_s$ which is dissipated as spintherm $\mathcal{Q}_s$ in the spin reservoir. The spin reservoir has its own inverse spin temperature $\gamma$. \label{fig:SHE}}
\end{figure}

This situation is radically changed when multiple conserved quantities are considered. In this case, since one does not need to rely on energy to erase the memory~\cite{Vaccaro2006,Vaccaro2009a,Vaccaro2011,Barnett2013}, it is therefore possible to design an engine where the output work amounts to \emph{all} of the heat extracted from a single thermal reservoir, as shown in Fig.~\ref{fig:SHE}. It must be noted that there is no ``free lunch'' and that the cost of dumping the entropy from the system still needs to be paid. It is just that, under this new paradigm, the cost can be paid using a different ``currency'', chosen from whatever other conserved quantity is available for use. Modern information-theoretical approaches to thermodynamics recognise this possibility, with a number of contributions describing the interplay between multiple conserved quantities in the cost of thermodynamical transformations~\cite{Guryanova2016,Ito2018,Bera2017}. 

In the situation considered in~\cite{Vaccaro2006,Vaccaro2009a,Vaccaro2011,Barnett2013}, the erasure cost is given specifically in terms of angular momentum. A generalised heat engine based on this scenario is depicted in Fig.~\ref{fig:SHE}, where the cold thermal bath is replaced by a polarized spin angular momentum reservoir. Following the work of Vaccaro and Barnett, we call this conceptual engine the VB spin-heat engine (SHE). The thermodynamical transformations in this engine take into account not only heat and work, but also the exchange of the previously defined {\it spinlabor} ${\cal L}_s$ and {\it spintherm} ${\cal Q}_s$. In contrast to the Carnot engine where the increase of entropy happens within the cooler thermal reservoir by heat flow, the SHE's entropy increases as a result of information erasure by dissipating an amount of spinlabor as spintherm into the spin reservoir, as shown in Fig.~\ref{fig:SHE}. Overall, the thermal reservoir cools down, and the spin reservoir experiences a reduction of polarisation to extract work.

Wright \emph{et al.} \cite{Wright2018} have recently designed an implementation of this generalised heat engine in a quantum dot system. The working fluid is represented by an electron trapped inside the quantum dot, which has an electronic level structure in a $\Lambda$ configuration with two energy degenerate ground states. 
The thermal and spin reservoirs in the diagram of Fig.~\ref{fig:SHE} are represented, respectively, by the phonons in the crystal and by the nuclear spin of the nuclei surrounding the confined electron.  

This quantum dot SHE works in three distinct stages. First, the electron is initially prepared in the ground state with spin up ($g_{\uparrow}$). Then a red-detuned Raman pulse is applied such that thermal energy $Q$ from the phonon reservoir is required for the transition to the spin down ground state ($g_{\dn}$) to occur, resulting in heat being converted into work $W$ in the form of coherent light. Note that in this process $W=Q$ but there is also an amount of spinlabour ${\cal L}_s=-\hbar$ provided to the dot by the laser pulse. A successful work extraction stage is marked by the transfer of electronic population from $g_{\uparrow}$ to  $g_{\downarrow}$. For the cycle to be complete, this information need to be erased, and the population reset to the original ground state $g_{\uparrow}$. This is achieved through the hyperfine interaction between the electron and the nuclei spins as described in Section~\ref{sec:discrete variables}. Since the two ground states are energy degenerate, the cost of information erasure in this last process is solely given in terms of spintherm, ${\cal Q}_s$, corresponding to the decrease in nuclear spin polarisation.

\section{Fluctuations}\label{sec:fluctuations}

The expressions of the cost of erasure in \eqs{eq:general cost of erasure of 1 bit}, \eqref{eq:erasure cost Q_1} and \eqref{eq:spin erasure} are in terms of average values taken over many experimental runs.  Each average value has an associated probability distribution with a finite spread, and so each run can result in a outcome that differs from the average value.
It is possible, therefore, that the cost of erasure in one run of an experiment can \emph{violate} the expressions given above.
Single values such as the standard deviation have long been used as measures of the extent any particular outcome may differ from the average.
However, the last three decades has seen the development of methods for studying the detailed statistics of the fluctuations in the outcomes over many runs \cite{Evans1993,Evans1994}.

For example, Dillenschneider and Lutz \cite{Dillenschneider2009} found the probability that the heat $Q$ dissipated in the reservoir in erasing 1 bit in one run of an experiment will violate Landauer's bound by the amount $\epsilon$ is given by
\begin{align}	
	P \left(Q< \beta^{-1}\ln 2 - \epsilon \right) < \exp(-\epsilon\beta)\ ,\label{eq:prob violation}
\end{align}
that is, the probability of violation falls off exponentially with $\epsilon$ and the inverse temperature $\beta=1/k_BT$.
Another way to say this is that large fluctuations are exponentially suppressed.

The corresponding Jarzynski equality \cite{Jarzynski1997} for the work $W$ performed in the erasure is 
\begin{align} \label{eq:Jarzynski}
	\left\langle \exp(-\beta W -\Delta F)\right\rangle =1
\end{align}
where $\Delta F=-\beta^{-1}\ln(Z_{f}/Z_{i})$ is the change in free energy and $Z_{i}$ and $Z_{f}$ are the initial and final partition functions. Note that $\Delta F$ has just one particular value over all experimental runs whereas the work $W$ fluctuates from one run to the next. 
In the case of the erasure of 1 bit of information $\Delta F=\beta^{-1}\ln2$.
Using Jensen's equality \cite{Jensen1906} with \eqref{eq:Jarzynski} yields
\begin{align}
	\left\langle W \right\rangle \geqslant \beta^{-1}\ln2
\end{align}
which shows that the bound is satisfied on average, as expected. 
These results assume that the erasure is carried out quasi-statically, i.e. very slowly, and have been verified experimentally in an ion trap using a Ca$^{+}$ ion by Xion \emph{et al.} \cite{Xiong2018}.
An experimental investigation of the fluctuations as the rate of the erasure process increases has also been reported by B\'{e}rut, Petrosyan and Ciliberto \cite{Berut2013} using a silica bead in an optical trap.

In contrast to the continuous variables treated above, we have analysed the fluctuations in the spinlabor cost $\mathcal{L}_{s}$ in the discrete case \cite{Croucher2017}.
The problem involved erasing of 1 bit of information stored in a spin-$\frac{1}{2}$ memory using a spin reservoir and the increasing gap erasure process described in Section~\ref{sec:discrete variables}.  The variable $\hat{V}_k$ in this case is $\hat{J}_z$, the $z$ component of angular momentum, and the bound on the spinlabor for the particular problem studied is
\begin{align}  \label{eq:spinlabor bound}
   \mathcal{L}_{s} \ge \frac{\ln 2}{\gamma} \ .
\end{align}
We found the fluctuations in $\mathcal{L}_{s}$ are described by the Jarzynski-like equality:
\begin{align}
	\left\langle \exp(-\gamma\mathcal{L}_{s} +\ln 2)\right\rangle =\frac{1+e^{-\gamma\hbar}}{1+e^{-2\gamma\hbar}}=A\ .\label{eq:Jarzynski like}
\end{align}
The non-unity value on the right side reflects the fact that the initial state of the memory, being maximally mixed, corresponds to a zero inverse spin temperature and is, therefore, out of equilibrium with the spin reservoir whose inverse spin temperature, $\gamma$, is finite.  This initial disequilibrium represents another feature that can occur when information is encoded using discrete observables.
Using  \eq{eq:Jarzynski like} and standard analytical techniques, 
we find the probability that the spinlabor cost will violate the bound in \eq{eq:spinlabor bound} by $\epsilon$ is given by
\begin{align}
P(\mathcal{L}_{s}\leq \gamma^{-1}\ln{2}-\epsilon ) \leq A \exp(-\gamma\epsilon)\label{eq:bound}
\end{align}
which, apart from the factor $A$, has the same form as \eq{eq:prob violation}. However, using  a semi-analytic method, we also found a tighter bound for $\hbar\gamma <1$ is given by
\begin{align}
     P(\mathcal{L}_{s}\leq \gamma^{-1}\ln{2}-\epsilon )\leq  C \exp(-\mbox{$\sqrt{\frac{\gamma}{\hbar}}$}\epsilon)
     \label{eq:tighter bound}\ ,
\end{align}
where $C=P(\mathcal{L}_{s}\leq \gamma^{-1}\ln{2})$.
That is, in the high spin-temperature limit (i.e. $\gamma^{-1}\gg \hbar$), large fluctuations are exponentially suppressed to a greater degree in comparison to the corresponding energy case in \eq{eq:prob violation}.
In this regime, the spins in the reservoir approach their maximally mixed state, and evidently this constrains the fluctuations.

\section{Discussion}\label{sec:discussion}

The history of the link between information and thermodynamics is a long and interesting one. Some
of the key steps were the discussion by Maxwell of his demon \cite{Maxwell1902}, Szilard's observation that acquiring
information could be associated with an entropy reduction \cite{Szilard1929} and, of course, Shannon's theory of
communications, which definitively determined that entropy is the mathematical measure of information  \cite{Shannon1963}.  Our topic in this chapter has been information erasure, which Landauer famously linked with a
minimum energy cost using thermodynamic principles, much as Szilard had done \cite{Landauer1961}. We have shown,
however, that the energy cost is not fundamental and that the cost of erasure can be paid by expending
other resources in place of energy \cite{Vaccaro2006,Vaccaro2009a,Vaccaro2011,Barnett2013,Lostaglio2017}.  Yet, the link between information erasure and
thermodynamics raises interesting questions for the second law of thermodynamics, although we must
be careful to keep in mind Eddington's warning \textit{``But if your theory is found to be against the second law of thermodynamics I can give you
no hope; there is nothing for it but to collapse in deepest humiliation.''} \cite{Eddington1928}

The strength of the link between Landauer's erasure principle and the second law was made clear by
Schumacher who showed that defeating Landauer's principle would enable us to break the second
law and also that if we can violate the second law then we can also erase information without incurring
the energy cost demanded by Landauer's principle \cite{Schumacher2011}. We have seen that it is possible to
erase information without incurring the energy cost and this necessarily implies a modification of the
second law of thermodynamics and, indeed, points to the necessity of modifying it by replacing the
familiar thermodynamical entropy by the more general expression \eqref{eq:S for general reservoir}.  Thus, logically, we may
trace the origins of generalisations of the second law \cite{Song2007,Brandao2015} back to information erasure
without an energy cost.  Perhaps we might even adopt the entropic cost of information erasure as a
physical principle.

\bigskip

\acknowledgments

\noindent
ACKNOWLEDGEMENTS

J.A.V. thanks the Australian Research Council (LP140100797) and the Lockheed Martin Corporation for financial support.  S.M.B. thanks the Royal Society for support (RP150122).

\bibliography{QuTherm}

\end{document}